\documentclass[fleqn,10pt]{wlscirep}
\usepackage[utf8]{inputenc}
\usepackage[T1]{fontenc}
\usepackage{enumitem}
\usepackage{amsmath}
\usepackage{lineno}
\usepackage{hyperref}       
\usepackage{url}            
\usepackage{booktabs}       
\usepackage{amsfonts}       
\usepackage{nicefrac}       
\usepackage{microtype}      
\usepackage{xcolor}         
\usepackage{soul}
\usepackage{mdframed}
\usepackage{verbatim}
\usepackage{listings}
\usepackage{pstricks}
\usepackage{pmboxdraw}
\usepackage{multirow}
\usepackage{graphicx}
\usepackage{array}
\usepackage{forest}
\usepackage{tikz}
\usepackage{fancyvrb}
\usepackage{caption}
\usepackage{geometry}
\usepackage{alltt}
\usepackage{xfakebold}

\definecolor{gray}{rgb}{0.4,0.4,0.4}
\definecolor{darkblue}{rgb}{0.0,0.0,0.6}
\definecolor{cyan}{rgb}{0.0,0.6,0.6}
\definecolor{backcolour}{rgb}{0.95,0.95,0.92}
\definecolor{OceanBlue}{HTML}{0d7377}
\definecolor{SteelBlue}{RGB}{70,130,180}

\AtBeginDocument{%
  }

\title{
MLOmics: Cancer Multi-Omics Database for Machine Learning
}

\author[1,2,$\dag$]{Ziwei Yang}
\author[2,$\dag$]{Rikuto Kotoge}
\author[2]{Xihao Piao}
\author[2,*]{Zheng Chen}
\author[3]{Lingwei Zhu}
\author[4]{Peng Gao}
\author[2]{Yasuko Matsubara}
\author[2]{Yasushi Sakurai}
\author[5,6]{Jimeng Sun}

\affil[1] {Bioinformatics Center, Institute for Chemical Research, Kyoto University, Japan}
\affil[2] {SANKEN, Osaka University, Japan}
\affil[3] {IRCN, The University of Tokyo, Japan}
\affil[4] {Institute for Quantitative Biosciences, The University of Tokyo, Japan}
\affil[5] {Department of Computer Science, University of Illinois Urbana-Champaign, USA}
\affil[6] {Carle Illinois College of Medicine, University of Illinois Urbana-Champaign, USA}
\affil[$\dag$] {These authors contributed equally to this work}
\affil[*]{Corresponding author}

\begin{abstract}

Framing the investigation of diverse cancers as a machine learning problem has recently shown significant potential in multi-omics analysis and cancer research. 
Empowering these successful machine learning models are the high-quality training datasets with sufficient data volume and adequate preprocessing. 
However, while there exist several public data portals, including The Cancer Genome Atlas (TCGA) multi-omics initiative or open-bases such as the LinkedOmics, these databases are not off-the-shelf for existing machine learning models. 
In this paper, we introduce MLOmics, an open cancer multi-omics database aiming at serving better the development and evaluation of bioinformatics and machine learning models.
MLOmics contains 8,314 patient samples covering all 32 cancer types with four omics types, stratified features, and extensive baselines. 
Complementary support for downstream analysis and bio-knowledge linking are also included to support interdisciplinary analysis.

\end{abstract}
\begin{document}

\flushbottom
\maketitle

\thispagestyle{empty}


\section*{Background \& Summary}




Multi-omics analysis has shown great potential to accelerate cancer research. 
A promising trend consists of framing the investigation of diverse cancers as a machine learning problem,
where complex molecular interactions and dysregulations associated with specific tumor cohorts are revealed through integration of multi-omics data into machine learning models. 
Several impressive achievements have been demonstrated in molecular subtyping \cite{MLomics2, guoyikebriefing, omicsECML}, disease-gene association prediction \cite{gysi2018wto, wu2021gaerf, genediseasenetworking1}, and drug discovery \cite{NatureDrug}.

Empowering successful machine learning models are the high-quality training datasets with sufficient data volume and adequate preprocessing.
While there exist several public data portals including The Cancer Genome Atlas (TCGA) multi-omics initiative \cite{TCGA} or open-bases such as the LinkedOmics \cite{LinkedOmics}, these databases are not off-the-shelf for existing machine learning models.
To make these data model-ready, a series of laborious, task-specific processing steps such as metadata review, sample linking, and data cleaning are mandatory.
The domain knowledge required, as well as a deep understanding of diverse medical data types and proficiency in bioinformatics tools have become an obstacle for researchers outside of such backgrounds.
The gap between the growing body of powerful machine learning models and the absence of well-prepared public data has become a major bottleneck.
Currently, some existing methods validate their proposed machine learning models using inconsistent experimental protocols, with variations in datasets, data processing techniques and evaluation strategies \cite{MLomics}.
These studies could benefit from a fair assessment of extensive baselines on a uniform footing with unified datasets { and to offer more reliable recommendations on models.}

\begin{figure}[p]
  \centering
  \includegraphics[width=\linewidth]{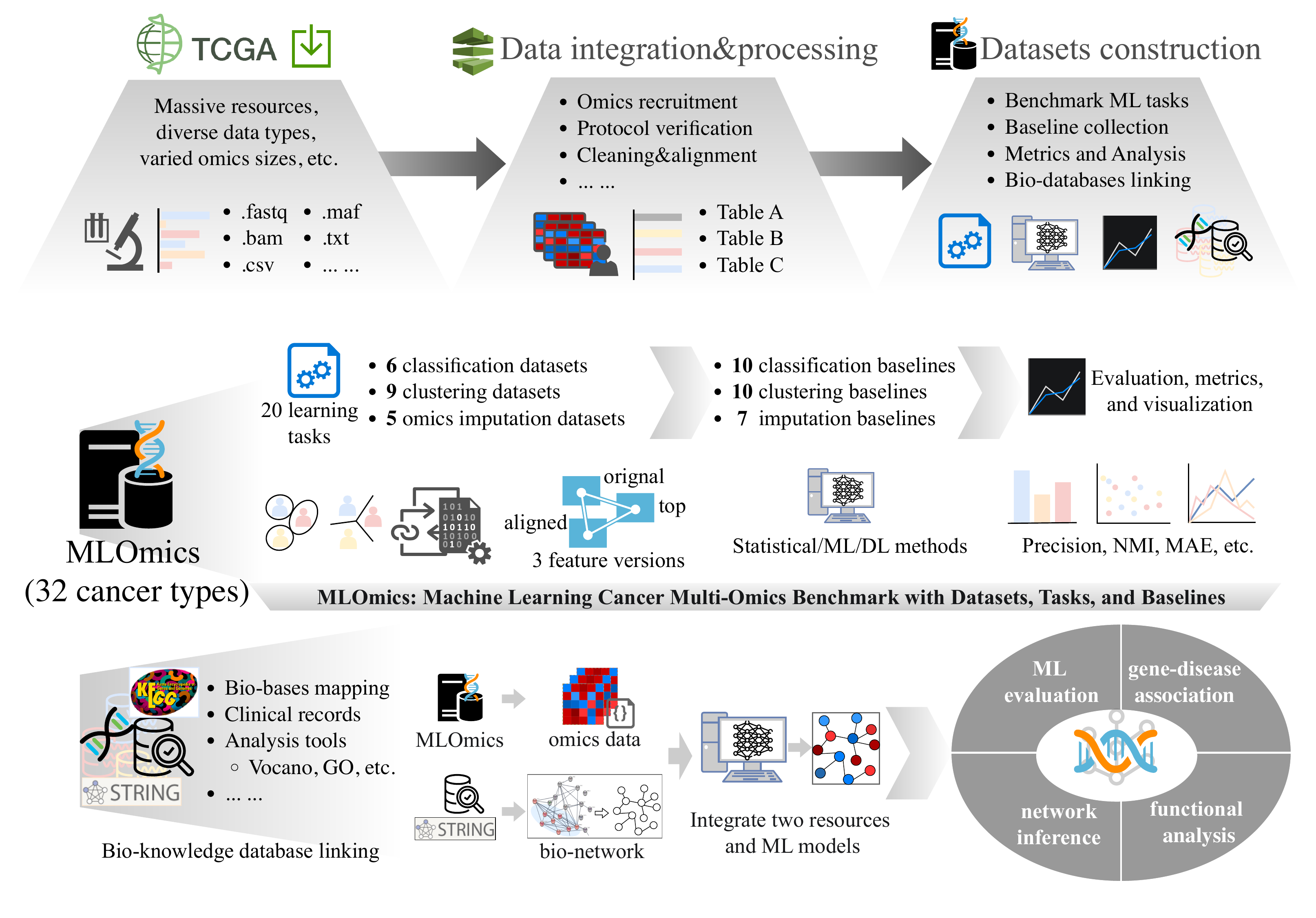}
  \caption{\textbf{Schematic workflow of creating the MLOmics.} The process starts with collecting patient samples covering 32 cancer types from the TCGA project. All resources in diverse data types and sizes are uniformly integrated and processed to contain data of four omics types. Datasets for benchmark ML tasks were constructed based on the processed data. MLOmics also selected baselines, metrics, and resources to support downstream biological analysis.
  \textbf{Overview of the MLOmics.} MLOmics provides an interface for developing and evaluating machine learning models based on cancer multi-omics data. MLOmics provides datasets in three feature scales for 20 classification, clustering, and omics imputation learning tasks. MLOmics also provides statistical, ML, and DL baselines for each task, which are evaluated by fair metrics.
  \textbf{Bio-knowledge database linking with MLOmics.} MLOmics provides resources to link with other bio-knowledge databases, enabling the integration of outer resources for applications such as ML evaluation, gene-disease association exploration, network inference, and functional analysis.}
  \label{fig:OV}
\end{figure}

To meet the growing demand of the community, { we introduce MLOmics, an open cancer multi-omics database} aiming at serving better the development and evaluation of bioinformatics and machine learning models.
MLOmics collected 8,314 patient samples covering all 32 cancer types from the TCGA project.
All samples were uniformly processed to contain four omics types: mRNA expression, microRNA expression, DNA methylation, and copy number variations, followed by categorization, protocol verification, feature profiling, transformation, and annotation.
For each dataset, we provide three feature versions: \texttt{Original}, \texttt{Aligned}, and \texttt{Top}, to support feasible analysis.
For example, the \texttt{Top} version contains the most significant features selected via the ANOVA test \cite{ANOVA} across all samples to filter out potentially noisy genes.
The MLOmics datasets were carefully examined with 6$\sim$10 highly cited baseline methods.
These baselines were rigorously reproduced and evaluated with various metrics to ensure fair comparison.
Complementary resources are included to support basic downstream biological analysis, such as clustering visualization, survival analysis, and Volcano plots. 
Last but not least, we provide support to interdisciplinary analysis via our locally deployed bio-base resources.
Interdisciplinary researchers can retrieve and integrate bio-knowledge of cancer omics studies through resources such as the STRING \cite{string} and KEGG \cite{kegg}.
For instance, exploration of bio-network inference \cite{bionetworkinference} and simulated gene knockouts \cite{knockout} is supported.
{In summary, MLOmics is an open, unified database approachable to non-experts for developing/evaluating machine learning models; conducting interdisciplinary analysis and supporting cancer research and broader biological studies.
We showcase the above usages in the Usage Notes section.
A detailed overview of the MLOmics database and its characteristics is provided in Figure \ref{fig:OV}.
}

\section*{Methods}
\label{sec: Methods}

\subsection*{Data Collection and Preprocessing}
The data were sourced from TCGA via the Genomic Data Commons (GDC) Data Portal \cite{GDC}.
The original resources in TCGA are organized by cancer type, and the omics data for individual patients are scattered across multiple repositories.
Therefore, retrieving and collecting omics data require linking samples with metadata and applying different preprocessing protocols.
MLOmics employs a unified pipeline that integrates data preprocessing, quality control, and multi-omics assembly for each patient, followed by alignment with their respective cancer types.
Specifically, we perform different steps for each omics type as follows:

\textbf{Transcriptomics (mRNA and miRNA) data: }
1. \emph{Identifying transcriptomics}. 
  We trace the downloaded data using the "experimental\_strategy" field in the metadata, marked as "mRNA-Seq" or "miRNA-Seq", then we verify that "data\_category" is labeled as "Transcriptome Profiling."
  2. \emph{Determining experimental platform}.
    We identify the experimental platform from the metadata, such as "platform: Illumina" or "workflow\_type: BCGSC miRNA Profiling."
  3. \emph{Converting gene-level estimates}.
    For data generated by platforms like Illumina Hi-Seq, use the edgeR package \cite{edger} to convert scaled gene-level RSEM estimates into FPKM values.
  4. \emph{Non-Human miRNA filtering}.
    For "miRNA-Seq" data from platforms like Illumina GA and Agilent arrays, we identify and remove non-human miRNA expressions using species annotations from databases such as miRBase \cite{miRBase}.
    5. \emph{Noise eliminating}.
    We remove features with zero expression in more than 10\% of samples or those with undefined value (N/A). 
    6. \emph{Transformation}. 
      Finally, we apply logarithmic transformations to obtain the log-converted mRNA and miRNA data.

\textbf{Genomic (CNV) data:}
1. \emph{Identifying CNV Alterations}.
We examine how gene copy-number alterations are recorded in the metadata, using key descriptions such as "Calls made after normal contamination correction and CNV removal using thresholds."
2. \emph{Filter Somatic Mutations}.
We capture only somatic variants by retaining entries marked as "somatic"  and filtering out germline mutations.
3. \emph{Identify Recurrent Alterations}.
We use the GAIA package \cite{gaia} to identify recurrent genomic alterations in the cancer genome, based on raw data representing all aberrant regions from copy-number variation segmentation.
4.  \emph{Annotate Genomic Regions}.
We annotate the recurrent aberrant genomic regions using the BiomaRt package \cite{BiomaRt}.

\textbf{Epigenomic (Methy) data:}
1. \emph{Identify Methylation Regions}.
We examine how methylation is defined in metadata to map methylation regions to genes, using key descriptions like "Average methylation (beta-values) of promoters defined as 500bp upstream \& 50 downstream of Transcription Start Site (TSS)" or "With coverage >= 20 in 70\% of the tumor samples"
2. \emph{Normalize Methylation Data.}
We perform median-centering normalization to adjust for systematic biases and technical variations across samples, using the R package limma \cite{limma}.
3. \emph{Select Promoters with Minimum Methylation}.
For genes with multiple promoters, we select the promoter with the lowest methylation levels in normal tissues.

After processing the omics sources, the data are annotated with unified gene IDs to resolve variations in naming conventions caused by the difference in sequencing methods or reference standards \cite{TCGAreview}.
Then, the omics data are aligned across multiple sources based on their corresponding sample IDs.
Finally, the data files are organized by cancer type for further dataset construction.

\subsection*{Datasets Construction}

{\color{black}
MLOmics reorganizes the collected and processed data resources into different feature versions tailored to various machine learning tasks. 
For each task, MLOmics provides several baselines, evaluation metrics, and the ability to link with biological databases such as STRING and KEGG for further biological analysis of different machine learning models.}

\label{subsec:datasetsconstruction}
\subsubsection*{Feature Processing.}
Machine learning models require tabular data with a the same number of features across samples.
In addition to the \texttt{Original} feature scale that contains a full set of genes (variations included) directly extracted from the collected omics files, MLOmics provides two other well-processed feature scales: \texttt{Aligned} and \texttt{Top}.
The former scale filters out non-overlapping genes and selects the genes shared across different cancer types; and the latter identifies the most significant features.
Specifically, the following steps are performed for each scale:

\textbf{Aligned:}
1. we resolve the mismatches in gene naming formats such as ensuring compatibility between cancers that use different reference genomes.
2.  we identify the intersection of feature lists across datasets to ensure all selected features are present in different cancers.
3.  we conduct z-score feature normalization.

\textbf{Top:} 
1.  we perform multi-class ANOVA \cite{ANOVA} to identify genes with significant variance across multiple cancer types.
2.   we perform multiple testing using the Benjamini-Hochberg (BH) \cite{bh} correction to control the false discovery rate (FDR) \cite{fdr}.
3.  the features are ranked by the adjusted $p$-values \(p < 0.05\) or by the user-specified scales).
4. we conduct z-score feature normalization which reduces the presence of non-significant genes across cancers and this could be beneficial for biomarker studies.

\subsubsection*{20 Task-ready Datasets with Baselines and Metrics.}

We provide 20 off-the-shelf datasets ready for machine learning models ranging from pan-cancer/cancer subtype classification, subtype clustering to omics data imputation.
We also include well recognized baselines that leveraged classical statistical approaches and machine/deep learning methods as well as metrics for standard evaluation.

\vspace{1mm}
\textbf{Pan-cancer and golden-standard cancer subtype classification. }
Pan-cancer classification aims to identify each patient's specific cancer type.
Moreover, a cancer typically comprises multiple subtypes that differ in their biochemical profiles.
Some subtypes have been well-studied in prior research and widely accepted as the golden standard.
We re-label patient samples to support subtyping evaluation.
These two classification tasks potentially improve cancer early diagnostic accuracy and treatment outcomes.

\emph{Datasets: } 
MLOmics provides six labeled datasets: one pan-cancer dataset and five gold-standard subtype datasets (GS-COAD, GS-BRCA, GS-GBM, GS-LGG, and GS-OV).

\emph{Baselines: }
  we opt for the following classical classification methods as baselines:
XGBoost \cite{XGBoost},
Support Vector Machines (SVM) \cite{SVM},
Random Forest (RF) \cite{RF}, and
Logistic Regression (LR) \cite{LR}.
Additionally, we include six popular, open-sourced deep learning methods:
Subtype-GAN \cite{GAN-subtype},
DCAP \cite{DCAP},
XOmiVAE \cite{XOmiVAE},
CustOmics \cite{CustOmics}, and
DeepCC \cite{DeepCC}.

\emph{Metrics:}
For classification evaluation, we opt for precision (Pre), recall (Re), and F1-score (F1).
Since clustering is the primary focus of the subtyping task, due to the limited sample size (typically <100), we provide normalized mutual information (NMI) and adjusted rand index (ARI) to evaluate the agreement between the clustering results obtained by different methods and the true labels.

\vspace{1mm}
\textbf{Cancer Subtype Clustering. } 
Cancer subtyping remains an open question under fierce debate for most cancers, especially rare types. 
Numerous studies propose various clustering methods to identify distinct groups by identifying different clusters to support downstream evaluation and discovery of new subtypes.

\emph{Datasets: }
MLOmics provides nine unlabeled rare cancer datasets (ACC, KIRP, KIRC, LIHC, LUAD, LUSC, PRAD, THCA, and THYM) for this learning task.

\emph{Baselines: }
In addition to the aforementioned Subtype-GAN, DCAP, MAUI, XOmiVAE, we also include six clustering methods:
Similarity Network Fusion (SNF) \cite{SNF},
Neighborhood-based Multi-Omics clustering (NEMO) \cite{NEMO},
Cancer Integration via Multi-kernel Learning (CIMLR) \cite{CIMLR},
iClusterBayes \cite{iClusterBayes},
moCluster \cite{mocluster}, and
MCluster-VAEs \cite{MCluster-VAEs}.

\emph{Metrics: }
To evaluate the goodness of clustering we opt for the classic Silhouette coefficient (SIL) \cite{sil} and log-rank test $p$-value on survival time (LPS) \cite{adjusted}.

\vspace{1mm}
\textbf{Omics Data Imputation. } 
In addition to classification and clustering, we also provide a data imputation task focusing on imputing multi-omics cancer data.
The collected omics data typically have missing values due to experimental limitations, technical errors, or inherent variability.
The imputation process is crucial for ensuring the integrity and usability of TCGA omics data \cite{you2020handling}.

\emph{Datasets: }
MLOmics provides five omics datasets with missing values (Imp-BRCA, Imp-COAD, Imp-GBM, Imp-LGG, and Imp-OV).
Given a full dataset as a matrix $D \in \mathbb{R}^{n \times m}$, we follow prior works  \cite{yoon2018gain, you2020handling} to generate a mask matrix $M \in \{0, 1\}^{n \times m}$ uniformly at random with the probability of missing $P(M_{ij} = 0) = r_{\text{miss}}$, and the probability of retaining $P(M_{ij} = 1) = 1 - r_{\text{miss}}$.
The final data matrix is obtained by multiplying element-wise the data matrix $D$ with the mask $M$. 
The missing level $ r_{\text{miss}} $ is selected from $[0.3, 0.5, 0.7]$.

\emph{Baselines: }
We opt for seven well-recognized methods for imputing missing values, including:
Mean Imputation that imputes the missing entry with mean values of entries around it (Mean);
K-Nearest Neighbors (KNN) that imputes the missing value with the weighted Euclidean K nearest neighbors;
Multivariate Imputation by Chained Equations (MICE) that performs multiple regression to model each missing value conditioned on non-missing values \cite{Buuren2011mice};
Iterative SVD (iSVD) that imputes by iterative low-rank SVD decomposition \cite{troyanskaya2001missing};
Spectral Regularization Algorithm (Spectral) that also employs SVD but with a soft threshold and the nuclear norm regularization \cite{mazumder2010spectral};
Generative Adversarial Imputation Nets (GAIN) that proposes to distinguish between fake and true missing patterns by the generator-discriminator architecture \cite{yoon2018gain};
Graph Neural Network for Tabular Data (GRAPE) that utilizes the graph networks to impute based on learned information from columns and rows of the data matrix \cite{you2020handling}.

\emph{Metrics: }
We use the Mean Squared Error (MSE) between the unmasked entries ($M_{ij}=1$) as the training loss to let the model predict the actual value.
During the test, the masked missing values are used for evaluating the model performance ($M_{ij} = 0$).

MLOmics will be continuously updated with baselines and evaluations on the defined learning tasks.
A detailed description of the baselines and metrics is provided in the Supplementary Material.

\section*{Data Records}
\label{sec: Records}
{

The MLOmics main datasets \cite{figshare} are available on Figshare (\url{https://figshare.com/articles/dataset/MLOmics_Cancer_Multi-Omics_Database_for_Machine_Learning/28729127}) and Hugging Face (\url{https://huggingface.co/datasets/AIBIC/MLOmics}).
The MLOmic main datasets are now accessible under the Creative Commons 4.0 Attribution (CC-BY-4.0), which supports its use for educational and research purposes. 
The main datasets presented include all cancer multi-omics datasets corresponding to the various tasks described above.
}
\subsubsection*{MLOmics Overview}
{
The MLOmics repository is structured into three sections: (1) Main Datasets, (2) Baseline and Metrics, and (3) Downstream Analysis Tools and Resources Linking.

The core section is the \textbf{(1) Main Datasets:} 
the repository hosts a comprehensive collection of tasks-ready cancer multi-omics datasets, stored primarily as .csv (comma-separated values) files.

Along with the other two sections:(2) Baseline and Metrics:
the repository provides source codes of baseline models and evaluation metrics for different tasks, typically implemented in Python or R code; 
(3) Downstream Analysis Tools and Resources Linking:
the repository encompasses additional tools and resources that complement the main datasets and downstream omics analysis needs, implemented as .csv, .py or R files.

In the following parts of this section, we focus on presenting the properties of the \textbf{Main Datasets}, detailing its organizational structure, principal components, and resources.

}
\subsubsection* {Main Datasets Format}

\begin{figure}[p]
  \includegraphics[width=\linewidth]{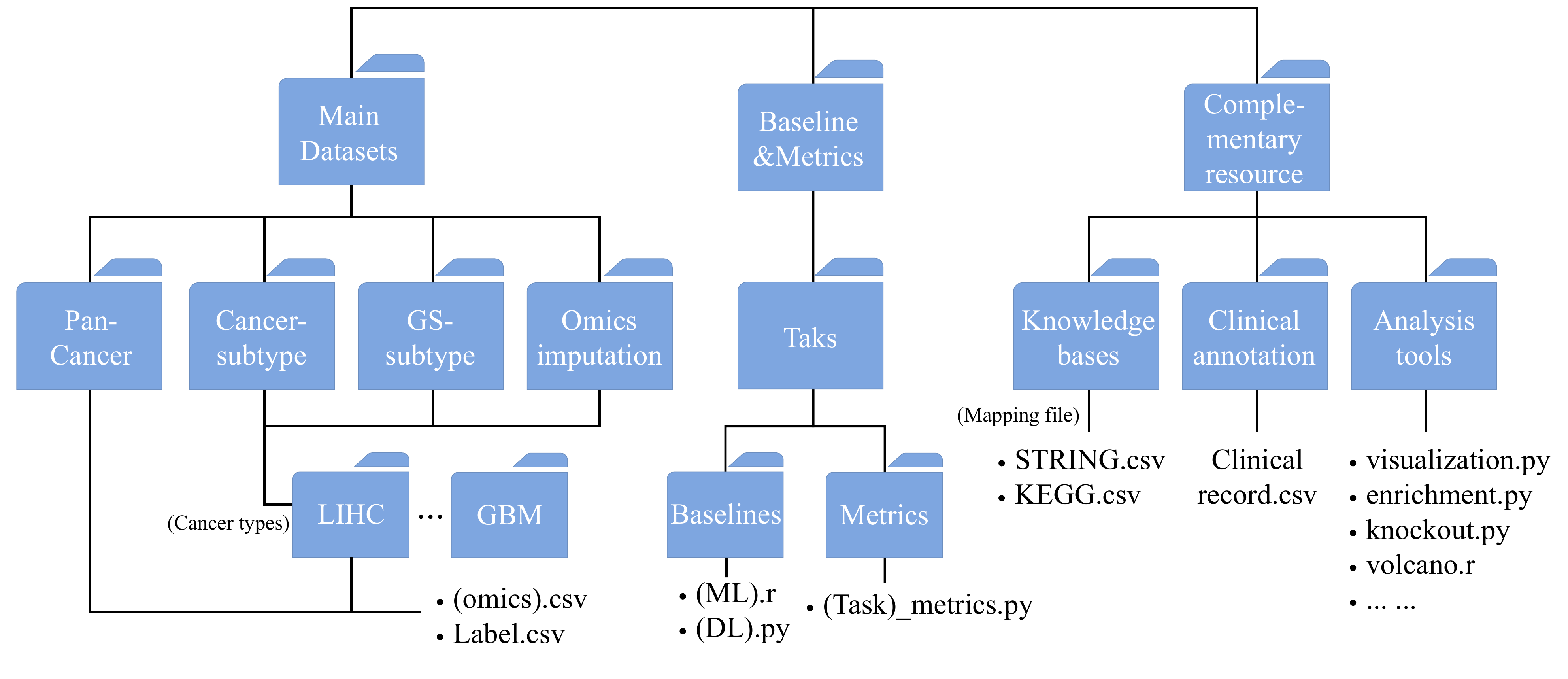}
  \caption{\textbf{Schematic of MLOmics resources structure.} The MLOmics framework consists of three major components.  The Main Datasets file includes all cancer multi-omics datasets for various tasks. The Baseline and Metrics file provides the source code for baseline models and performance metrics. The Downstream Analysis Tools and Resources Linking file offers sources for further analysis and links to additional biological resources.}
  \label{fig:folders}
\end{figure}

{

Main datasets of MLOmics use .csv files to manage and store all omics datasets, widely favored in biomedical research, including multi-omics studies.
Despite its simplicity, .csv remains efficient even with large datasets encountered in genomic and proteomic studies. 
Both Python and R provide built-in functions and libraries to read, write, and analyze .csv files efficiently.

Specifically, MLOmics provides its main datasets in .csv files as plain-text files, where commas separate data values.  
These files maintain a structured and intuitive format, with rows representing individual data records and columns corresponding to different attributes or variables.  
The key characteristics of the main dataset files are:  

\begin{itemize}[left=0pt]
\item \textbf{Encoding:} UTF-8, ensuring broad compatibility across different systems and software.
\item \textbf{Delimiter:} \texttt{,} (comma), used to separate values in each row.
\item \textbf{Line Ending:} LF (\texttt{\textbackslash n}) for Unix/Linux or CRLF (\texttt{\textbackslash r\textbackslash n}) for Windows, ensuring proper formatting across operating systems.
\item \textbf{Header Row:} The first row contains column names, typically representing sample identifiers (e.g., \texttt{Sample1}, \texttt{Sample2}, \texttt{Sample3}, etc.).
\item \textbf{Data Rows:} Each subsequent row corresponds to a distinct feature, such as a gene, with attributes like expression values recorded across multiple patient samples.
\item \textbf{Values:} Numeric values (\texttt{float}), representing continuous measurements of omics attributes.
\end{itemize}

For example, consider a simplified .csv file containing mRNA data for a set of gene features (rows) across several patient samples (columns):

\begin{center}
\begin{tabular}{lcccc}
\textbf{Feature} & \textbf{Sample1} & \textbf{Sample2} & \textbf{Sample3} & \textbf{Sample4} \\
\hline
GeneA & 0.23 & 0.18 & 0.35 & 0.21 \\
GeneB & 0.56 & 0.49 & 0.52 & 0.58 \\
GeneC & 0.19 & 0.22 & 0.15 & 0.17 \\
GeneD & 0.08 & 0.10 & 0.09 & 0.12 \\
\end{tabular}
\end{center}

In this example, each row corresponds to a specific gene (GeneA, GeneB, GeneC, GeneD).
Each column represents a different sample (Sample1, Sample2, Sample3, Sample4).
The numeric values (variables) in the cells denote the expression levels of each gene in each sample.

\subsubsection*{Main Datasets Structure}
As shown in Fig. \ref{fig:folders}, the main datasets repository is organized into three layers:

\vspace{1mm}
\textbf{(1) The first layer} contains three files corresponding to three different tasks: \texttt{Classification\_datasets}, \texttt{Clustering\_datasets}, and \texttt{Imputation\_datasets}.

\vspace{1mm}
\textbf{(2) The second layer} includes files for specific tasks, such as \texttt{GS-BRCA}, \texttt{ACC}, and \texttt{Imp-BRCA}.

\vspace{1mm}
\textbf{(3) The third layer} contains three files corresponding to different feature scales, i.e., \texttt{Original}, \texttt{Aligned}, and \texttt{Top}.

The omics data from different omics sources are stored in the following files: \texttt{Cancer\_mi\-RNA\_Feature.csv}, \texttt{Cancer\_mRNA\_Feature.csv}, \texttt{Cancer\_CNV\_Feature.csv}, and \texttt{Cancer\_Me\-thy\_Feature.csv}.
Here, \texttt{Cancer} represents the cancer type, and \texttt{Feature} indicates the feature scale type.  

The ground truth labels are provided in the file \texttt{Cancer\_label\_num.csv}, where \texttt{Cancer} represents the cancer type.  
The patient survival records are stored in the file \texttt{Cancer\_survival.csv}.

}

\section*{Technical Validation}
\label{sec: Experiments}
\begin{figure}[p]
  \centering
  \includegraphics[width=0.95\linewidth]{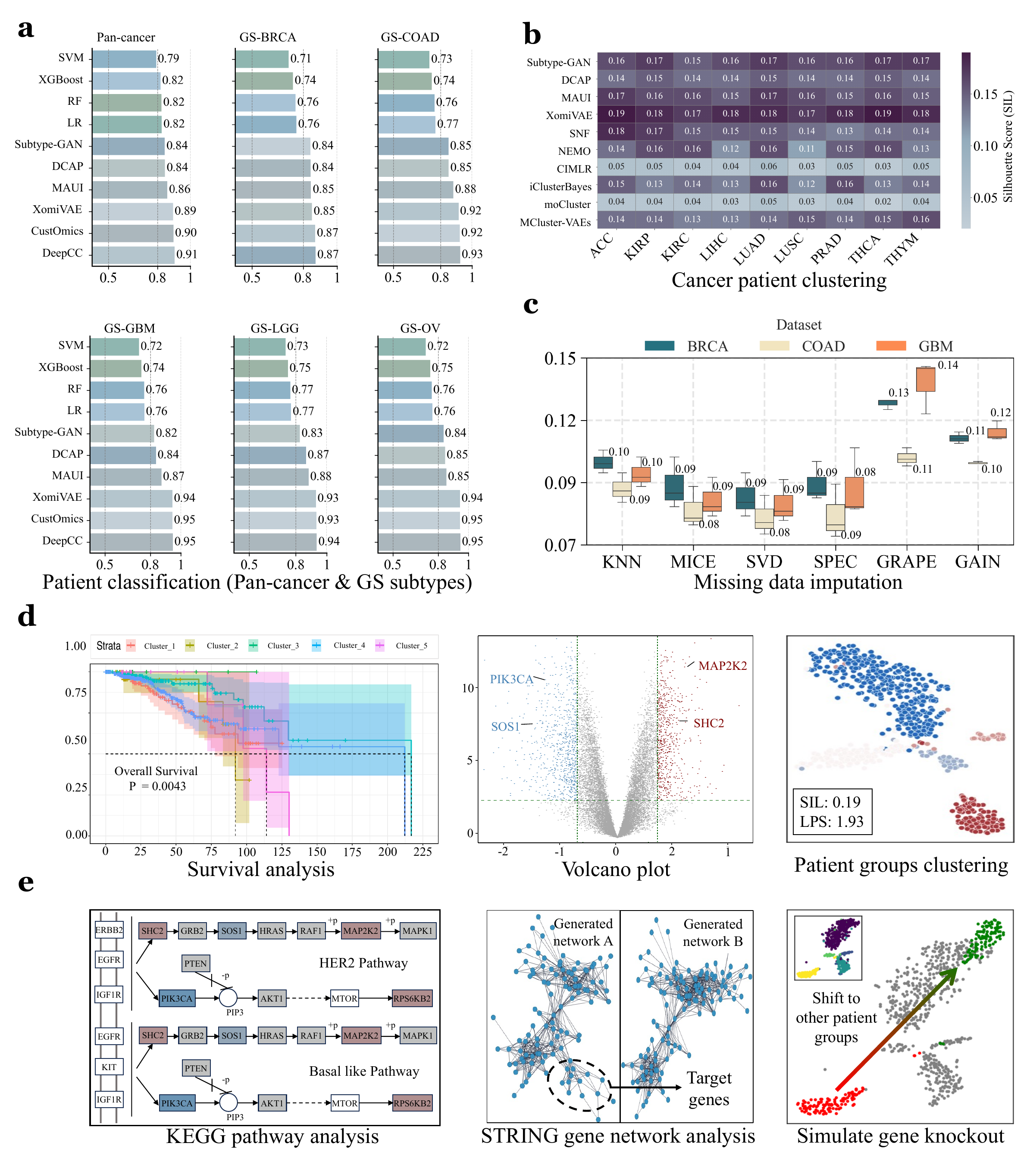}
  \caption{\textbf{Experimental results and downstream analyses of machine learning baselines applied to MLOmics datasets.} \textbf{(a)} PREC bar plots for each baseline method across all datasets. Overall, machine learning-based methods (Subtype-GAN, DCAP, MAUI, XOmiVAE, CustOmics, and DeepCC) outperformed traditional statistical methods (SVM, XGBoost, RF, and LR). \textbf{(b)} SIL heatmaps for each baseline method across all datasets. Methods employing deep generative neural network architectures (Subtype-GAN, DCAP, MAUI, XOmiVAE, and MCluster-VAEs) generally outperformed other methods (SNF, NEMO, CIMLR, iClusterBayes, and moCluster). \textbf{(c)} Box plots for each baseline method across three imputation datasets. Matrix decomposition methods (SVD, Spectral) outperformed deep learning-based methods (GAIN, GNN). \textbf{(d–e)} Schematic illustrations of downstream analysis results based on the clustering outcomes of XOmiVAE applied to specific cancer patient clustering datasets. In the survival analysis plot, survival curves in different colors correspond to distinct clustering groups. In the volcano plot and KEGG pathway analysis, red and blue indicate downregulated and upregulated genes between patient groups, respectively. In the patient group clustering plot, different colors represent samples belonging to different clusters. In the simulated gene knockout analysis, red and green dots indicate sample clustering before and after expression knockout, respectively.}
  \label{fig: Results}
\end{figure}

{
The Methods section detailed the data collection, preprocessing and curation of the MLOmics datasets.
To validate the collected datasets,  we set up a series of classification, clustering and imputation experiments each with a wide array of models ranging from conventional statistical models to deep neural networks. 
The experimental results and evaluations are summarized in Figure~\ref{fig: Results}.}

\section*{Usage Notes}
\begin{figure}[p]
  \centering
  \includegraphics[width=\linewidth]{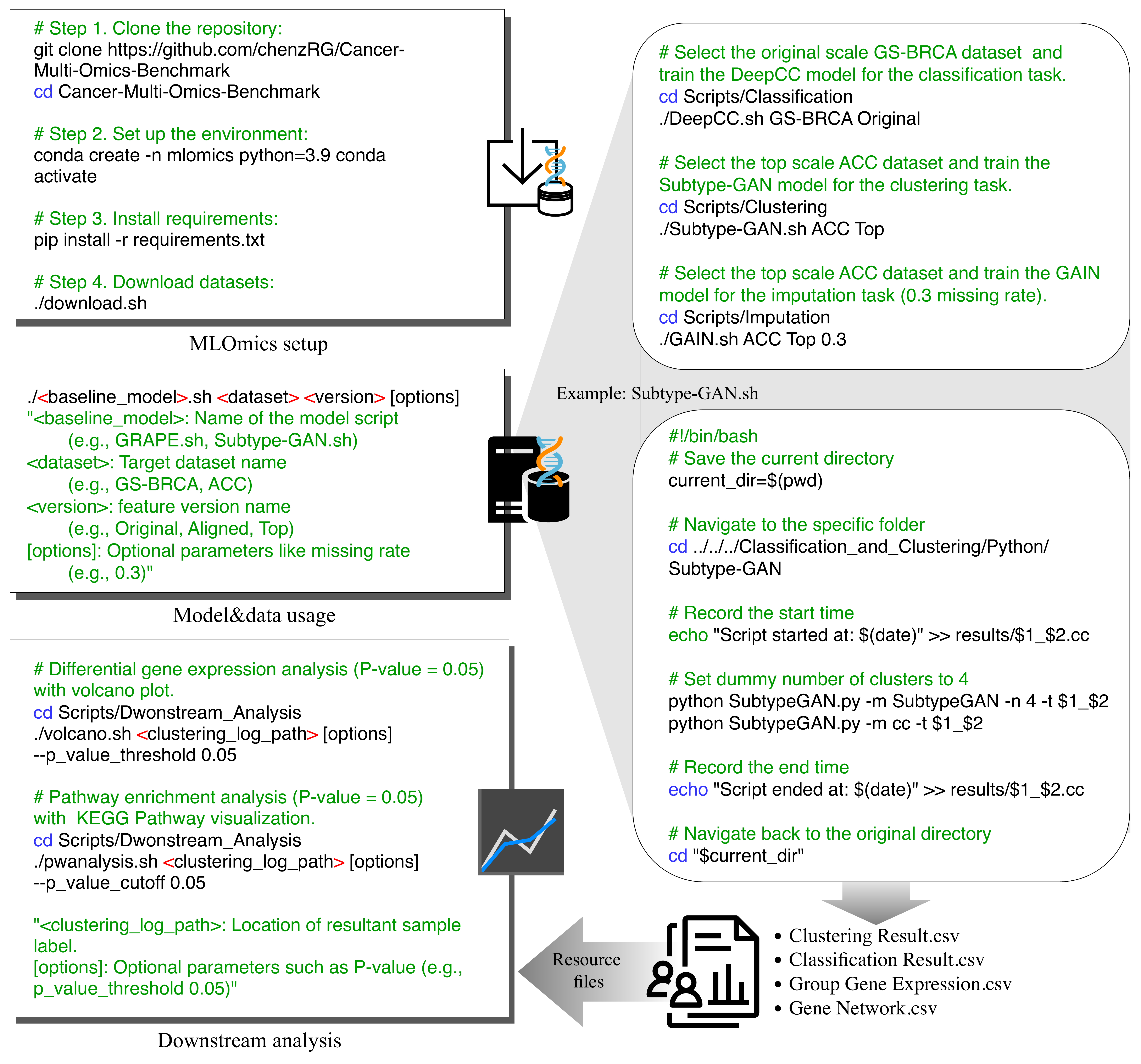}
  \caption{\textbf{The code usage note of MLOmics.} The MLOmics can be used following the illustrated code usage instructions, including MLOmics setup, code usage for custom model-task-dataset specifications, and code for downstream analysis workflows.}
  \label{fig:usage}
\end{figure}

We provide comprehensive guidelines for utilizing the MLOmics repository, as illustrated in Figure~\ref{fig:usage}.
All codes and files are ready for direct loading and analysis using standard Python data packages such as NumPy and Pandas.
{
As a demonstration of how to use MLOmics for machine learning applications, we provide 20 classification, clustering, and imputation tasks with fair evaluation protocols for pan-cancer analysis, cancer subtypes, and omics imputation.
Access to these resources is provided in the Code Availability section.}
A rising trend in multi-omics analysis is to integrate multi-omics data (non-network data) with biological networks to better understand complex functions on the gene or protein level.
MLOmics provides offline linking resources for well-established databases such as STRING \cite{string} and KEGG \cite{kegg}.
We hope it can lower the barriers to entry for machine learning researchers interested in developing methods for cancer multi-omics data analysis, thereby encouraging rapid progress in the field.


\section*{Code Availability}
{
All codes and resources of MLOmics are publicly available under the CC-BY-4.0. 
The code for preprocessing and preparing the main MLOmics databases, as well as the benchmarking algorithms used in MLOmics, is available at the following repository: (\url{https://github.com/chenzRG/Cancer-Multi-Omics-Benchmark}). 
}
\bibliography{reference.bib}

\section*{Acknowledgements}

This work was supported by
JSPS KAKENHI Grant-in-Aid for Scientific Research Number
JP24K20778,    
JST CREST JPMJCR23M3,
NSF award SCH-2205289, SCH-2014438, IIS-2034479.

\section*{Author contributions statement}

Zheng Chen and Ziwei Yang launched the project.
Ziwei Yang collected the data and organized the datasets.
Rikuto Kotoge built the dataset repository for storage and access.
Ziwei Yang, Rikuto Kotoge, and Xihao Piao conducted the experiments.
Zheng Chen, Lingwei Zhu, Ziwei Yang, and Peng Gao contributed to writing the manuscript and reviewed the article.
Peng Gao contributed to the biological review.
Zheng Chen, Yasuko Matsubara, Yasushi Sakurai, and Jimeng Sun conceptualized and supervised the project.

\section*{Competing interests} 

The authors declare no competing interests.


\clearpage
\clearpage

\clearpage
\clearpage

\clearpage
\clearpage

\clearpage
\clearpage

\thispagestyle{empty}
\begin{center}
  \Huge \textbf{Appendix} \\[1em]
\end{center}
\appendix
\section{Key Information about MLOmics}
\subsection{MLOmics Structure}
Here, we present the organizational structure of the MLOmics, detailing its main components and resources.
The MLOmics repository is structured into three primary sections: \textbf{Main Datasets}, \textbf{Baseline and Metrics}, and \textbf{Downstream Analysis Tools and Resources Linking}.

\begin{verbatim}
MLOmics
├── Main Datasets
│   ├── Pan-Cancer Dataset [Classification_datasets]
│   │   └── Pan-Cancer
│   │       └── Aligned
│   │           └── mRNA / miRNA / CNV / Methy / Label.csv
│   │
│   ├── Cancer Subtype Datasets [Clustering_datasets]
│   │   ├── ACC
│   │   ├── KIRP
│   │   ├── KIRC
│   │   ├── LIHC
│   │   ├── LUAD
│   │   ├── LUSC
│   │   ├── PRAD
│   │   ├── THCA
│   │   └── THYM
│   │   (ABOVE)└── Original & Aligned & Top
│   │              └── mRNA / miRNA / CNV / Methy.csv
│   │
│   ├── Golden-Standard Cancer Subtype Datasets [Classification_datasets]
│   │   ├── GS-COAD
│   │   ├── GS-BRCA
│   │   ├── GS-GBM
│   │   ├── GS-LGG
│   │   └── GS-OV
│   │   (ABOVE)└── Original & Aligned & Top
│   │              └── mRNA / miRNA / CNV / Methy / Label.csv
│   │
│   ├── Omics Data Imputation Datasets [Imputation_datasets]
│   │   ├── Imp-COAD
│   │   ├── Imp-BRCA
│   │   ├── Imp-GBM
│   │   ├── Imp-LGG
│   │   └── Imp-OV
│   │   (ABOVE)└── Top
│   │              └── mRNA / miRNA / CNV / Methy.csv
│   │
├── Baseline Models and Metrics
│   ├── Classification Tasks
│   │   ├── Baselines.py
│   │   └── Metrics.py
│   │
│   ├── Clustering Tasks
│   │   ├── Baselines.py/r 
│   │   └── Metrics.py 
│   │
│   └── Omics Data Imputation Tasks
│       ├── Baselines.py 
│       └── Metrics.py 
│
└── Downstream Analysis Tools and Resources Linking
    ├── Knowledge_bases
    │   └── STRING_mapping / KEGG_mapping.csv
    │   
    ├── Clinical Annotation
    │   └── Clinical_Rec.csv  
    │
    └── Analysis Tools
        └── Analysis_Tools.py/r  
\end{verbatim}






\subsection{Recruited Cancer}
The MLOmics database contains multi-omics data for 32 types of cancer.
The full names and abbreviations as shown in the following table:

\newpage
\begin{table}[h!]
\centering
\begin{tabular}{|c| m{7.5cm} | c |}
\hline
\textbf{No.} & \textbf{Full Name} & \textbf{Abbreviation} \\ \hline
1 & Acute Myeloid Leukemia & LAML \\ \hline
2 & Adrenocortical Cancer & ACC \\ \hline
3 & Bladder Urothelial Carcinoma & BLCA \\ \hline
4 & Brain Lower Grade Glioma & LGG \\ \hline
5 & Breast Invasive Carcinoma & BRCA \\ \hline
6 & Cervical \& Endocervical Cancer & CESC \\ \hline
7 & Cholangiocarcinoma & CHOL \\ \hline
8 & Colon Adenocarcinoma & COAD \\ \hline
9 & Diffuse Large B-cell Lymphoma & DLBC \\ \hline
10 & Esophageal Carcinoma & ESCA \\ \hline
11 & Head \& Neck Squamous Cell Carcinoma & HNSC \\ \hline
12 & Kidney Chromophobe & KICH \\ \hline
13 & Kidney Clear Cell Carcinoma & KIRC \\ \hline
14 & Kidney Papillary Cell Carcinoma & KIRP \\ \hline
15 & Liver Hepatocellular Carcinoma & LIHC \\ \hline
16 & Lung Adenocarcinoma & LUAD \\ \hline
17 & Lung Squamous Cell Carcinoma & LUSC \\ \hline
18 & Mesothelioma & MESO \\ \hline
19 & Ovarian Serous Cystadenocarcinoma & OV \\ \hline
20 & Pancreatic Adenocarcinoma & PAAD \\ \hline
21 & Pheochromocytoma \& Paraganglioma & PCPG \\ \hline
22 & Prostate Adenocarcinoma & PRAD \\ \hline
23 & Rectum Adenocarcinoma & READ \\ \hline
24 & Sarcoma & SARC \\ \hline
25 & Skin Cutaneous Melanoma & SKCM \\ \hline
26 & Stomach Adenocarcinoma & STAD \\ \hline
27 & Testicular Germ Cell Tumor & TGCT \\ \hline
28 & Thymoma & THYM \\ \hline
29 & Thyroid Carcinoma & THCA \\ \hline
30 & Uterine Carcinosarcoma & UCS \\ \hline
31 & Uterine Corpus Endometrioid Carcinoma & UCEC \\ \hline
32 & Uveal Melanoma & UVM \\ \hline
\end{tabular}
\caption{Cancer Types and Abbreviations in MLOmics}
\end{table}

\subsection{Recruited Omics}
MLOmics recruited four types of omics data:

\begin{itemize}[left=0pt]
\item
\emph{mRNA} (mRNA expression) measures the levels of messenger RNA transcribed from genes, reflecting the active transcription of genetic information; 
\item
\emph{miRNA} (miRNA expression) quantifies the levels of microRNAs and small non-coding RNA molecules. It is crucial for post-transcriptional regulation in gene expression;
\item
\emph{Methy} (DNA methylation) measures the addition of methyl groups to DNA, typically at cytosine bases. It influences gene expression by altering the DNA accessibility to transcriptional machinery.

\item
\emph{CNV} (copy number variations)  represents variations in the number of copies of particular DNA segments. This omics affects gene dosage and contributes to cancer susceptibility.
\end{itemize}

\newpage
\section{MLOmics Preprocessing Pipelines}
\label{SM: Pipelines}
Here are the details for processing different omics:
\subsection{For Transcriptomics (mRNA and miRNA) Data}
\begin{enumerate}
\item \textbf{STEP 1: Identify Transcriptomics Data}
Trace the data by "experimental\_strategy" in the metadata, marked as "mRNA-Seq" or "miRNA-Seq".
Check if "data\_category" is marked as "Transcriptome Profiling".
\item \textbf{STEP 2: Determine Experimental Platform}
Identify the experimental platform from metadata, such as "platform: Illumina" or "workflow\_type: BCGSC miRNA Profiling".
\item \textbf{STEP 3: Convert Gene-Level Estimates}
For data from the Hi-Seq platform like Illumina, use the R package edgeR \cite{edger} to convert the scaled estimates in the original gene-level RSEM to FPKM.
\item \textbf{STEP 4: Filter Non-Human miRNA}
For "miRNA-Seq" data from Illumina GA and Agilent array platforms, identify and remove non-human miRNA expression features using species annotation from databases like miRBase \cite{miRBase}.
\item \textbf{STEP 5: Eliminate Noise}
Identify and eliminate features with zero expression levels in more than 10\% of samples or missing values (designated as N/A).
\item \textbf{STEP 6: Apply Logarithmic Transformation}
Apply a logarithmic transformation to get the log-converted mRNA and miRNA data.
\end{enumerate}

\subsection{For Genomic (CNV) Data}
\begin{enumerate}
\item \textbf{STEP 1: Identify CNV Alterations in Metadata}
 Examine how alterations in gene copy-number are recorded in metadata using key descriptions like "Calls made after normal contamination correction and CNV removal using thresholds."
 \item \textbf{STEP 2: Filter Somatic Mutations}
 Use keyword filtering to capture only somatic mutations, excluding germline mutations by retaining only those marked as 'somatic."
 \item \textbf{STEP 3: Identify Recurrent Alterations}
Use the R package GAIA \cite{gaia} to identify recurrent alterations in the cancer genome from raw data that denote all aberrant regions resulting from copy number variation segmentation.
\item \textbf{STEP 4: Annotate Genomic Regions}
Use the R package BiomaRt \cite{BiomaRt} to annotate the aberrant recurrent genomic regions.
\item \textbf{STEP 5: Save Annotated CNV Data}
Save the annotated results to get CNV data of significantly amplified or deleted genes.
\end{enumerate}

\subsection{For Epigenomic (Methy) Data}
\begin{enumerate}
\item \textbf{STEP 1: Identify Methylation Regions in Metadata}
Examine how methylation is defined in metadata to map methylation regions to genes, using key descriptions like "Average methylation (beta-values) of promoters defined as 500bp upstream \& 50 downstream of Transcription Start Site (TSS)" or "With coverage >= 20 in 70\% of the tumor samples and 70\% of the normal samples."
\item \textbf{STEP 2: Normalize Methylation Data}
Implement a median-centering normalization to account for systematic biases and technical variations across samples using the R package limma\cite{limma}.
\item \textbf{STEP 3: Select Promoters with Minimum Methylation}
For genes with multiple promoters, select the promoter with minimum methylation in the normal tissues.  
\item \textbf{STEP 4: Save Mapped Methylation Data}
Save the mapped value data where each entry corresponds to a specific gene or genomic region, along with corresponding methylation measurements.
\end{enumerate}

\newpage
\section{MLOmics Feature Scale Processing}
\label{SM: Feature Scale}
Cancer multi-omics analysis often suffers from data issues, such as unbalanced sample sizes and feature dimensions. 

To address this, after the typical omics data preprocessing, MLOmics provides three versions of feature scales (\textit{Original}, \textit{Top}, and \textit{Aligned}) to support different machine learning tasks.
Table~\ref{tb: datasets} summarizes the different feature scales' details.

\subsection{Original Features}
The original features are genes directly extracted from each preprocessed omics dataset. 
It represents the complete, full-size set of patients' gene features without any task-specific filtering. 
This version supports users in customizing their datasets based on their specific requirements, such as re-filtering to target gene sets or omics-specific transformations.

Key technical operations in generating original features include:
\begin{enumerate}
\item Retaining all genes after normalization (e.g., log transformation or z-score normalization).
\item Imputing missing values using methods like K-nearest neighbors (KNN) or median imputation.
\item Filtering out low-quality samples (e.g., samples of low-variances or high missing genes).
\end{enumerate}

\subsection{Aligned Features}
Aligned features are the intersection of genes common to all datasets for a learning task, representing the shared features in different cancer types. 
This reduces the original feature size, ensures consistency in multi-omics features, and primarily provides support across cancer-type studies. 

Key technical operations for generating aligned features include:
\begin{enumerate}
\item Resolving unmatches in gene naming formats (e.g., ensuring compatibility between cancers using different references genome).
\item Identifying the intersection of feature lists across datasets to ensure all selected features are present in different cancers.
\item Normalization features(e.g., log transformation or z-score normalization).
\end{enumerate}

\subsection{Top Features}
Top features are selected based on ANOVA \cite{ANOVA} statistical testing, ranked by p-values to identify the most significant features across cancers. 
The default top feature scale settings for mRNA, miRNA, methylation, and CNV data are 5000, 200, 5000, and 5000, respectively, with significance determined by \(p < 0.05\). 
This approach greatly reduces noisy genes across cancers and achieves smaller feature-dimension cancer datasets, making them suitable for feature-dimension-sensitive machine-learning models.

Additional key technical operations for generating top features include:
\begin{enumerate}
\item Performing multi-class ANOVA to identify genes with significant variance across multiple cancer types.
\item Adjusting for multiple testing using the Benjamini-Hochberg correction to control the false discovery rate (FDR).
\item Ranking features by adjusted p-values and selecting the top \(k\) features per omics type, as defined by the default or user-specified scales.
\item Normalization features(e.g., log transformation or z-score normalization).
\end{enumerate}

The detailed feature size of different MLOmics datasets is below:

\begin{table}
\centering
\caption{MLOmics provides multiple feature scales for nine unlabeled cancer subtype datasets and five labeled, golden-standard subtype datasets.}
\label{tb: datasets}
\begin{tabular}{lccccc} 
\toprule
\multirow{2}{*}{Dataset} & \multirow{2}{*}{Feature Scale} & \multicolumn{4}{c}{Omics Feature Size}  \\ 
\cmidrule{3-6}
                         &                                & mRNA  & miRNA & Methy & CNV             \\ 
\midrule
\multirow{3}{*}{ACC}     & Orignal                        & 18204 & 368   & 19045 & 19525           \\
                         & Aligned                        & 10452 & 254   & 10347 & 10154           \\
                         & Top                            & 5000  & 200   & 5000  & 5000            \\
\multirow{3}{*}{KIRP}    & Orignal                        & 17254 & 375   & 19023 & 19532           \\
                         & Aligned                        & 10452 & 254   & 10347 & 10154           \\
                         & Top                            & 5000  & 200   & 5000  & 5000            \\
\multirow{3}{*}{KIRC}    & Orignal                        & 18464 & 352   & 19045 & 19523           \\
                         & Aligned                        & 10452 & 254   & 10347 & 10154           \\
                         & Top                            & 5000  & 200   & 5000  & 5000            \\
\multirow{3}{*}{LIHC}    & Orignal                        & 17945 & 435   & 19053 & 19523           \\
                         & Aligned                        & 10452 & 254   & 10347 & 10154           \\
                         & Top                            & 5000  & 200   & 5000  & 5000            \\
\multirow{3}{*}{LUAD}    & Orignal                        & 18303 & 435   & 19034 & 19532           \\
                         & Aligned                        & 10452 & 254   & 10347 & 10154           \\
                         & Top                            & 5000  & 200   & 5000  & 5000            \\
\multirow{3}{*}{LUSC}    & Orignal                        & 18577 & 745   & 19025 & 19543           \\
                         & Aligned                        & 10452 & 254   & 10347 & 10154           \\
                         & Top                            & 5000  & 200   & 5000  & 5000            \\
\multirow{3}{*}{PRAD}    & Orignal                        & 17954 & 467   & 19034 & 19534           \\
                         & Aligned                        & 10452 & 254   & 10347 & 10154           \\
                         & Top                            & 5000  & 200   & 5000  & 5000            \\
\multirow{3}{*}{THCA}    & Orignal                        & 17480 & 345   & 19024 & 19532           \\
                         & Aligned                        & 10452 & 254   & 10347 & 10154           \\
                         & Top                            & 5000  & 200   & 5000  & 5000            \\
\multirow{3}{*}{THYM}    & Orignal                        & 18341 & 535   & 19034 & 19532           \\
                         & Aligned                        & 10452 & 254   & 10347 & 10154           \\
                         & Top                            & 5000  & 200   & 5000  & 5000            \\ 
\midrule
\multirow{3}{*}{GS-COAD} & Orignal                        & 18234 & 462   & 19023 & 19545           \\
                         & Aligned                        & 11343 & 286   & 11189 & 11203           \\
                         & Top                            & 5000  & 200   & 5000  & 5000            \\
\multirow{3}{*}{GS-BRCA} & Orignal                        & 18233 & 345   & 19053 & 19533           \\
                         & Aligned                        & 11343 & 286   & 11189 & 11203           \\
                         & Top                            & 5000  & 200   & 5000  & 5000            \\
\multirow{3}{*}{GS-GBM}  & Orignal                        & 17545 & 335   & 19034 & 19545           \\
                         & Aligned                        & 11343 & 286   & 11189 & 11203           \\
                         & Top                            & 5000  & 200   & 5000  & 5000            \\
\multirow{3}{*}{GS-LGG}  & Orignal                        & 18345 & 345   & 19023 & 19534           \\
                         & Aligned                        & 11343 & 286   & 11189 & 11203           \\
                         & Top                            & 5000  & 200   & 5000  & 5000            \\
\multirow{3}{*}{GS-OV}   & Orignal                        & 1735  & 244   & 19034 & 19534           \\
                         & Aligned                        & 11343 & 286   & 11189 & 11203           \\
                         & Top                            & 5000  & 200   & 5000  & 5000            \\
\bottomrule
\end{tabular}
\end{table}

\newpage
\section{MLOmics Tasks}

The MLOmics database contains multi-omics data curated for a range of task categories, as summarized in Table~\ref{table: all_tasks}.
Definitions for each task category are provided below.

\begin{table}[t]
\centering
\caption{Summary of all tasks in the MLOmics under each task category.}
\label{table: all_tasks}
\resizebox{0.95\linewidth}{!}{%
\begin{tabular}{lll} 
\toprule
Entry & Task Categories                        & Task Names                                           \\ 
\midrule
1     & Pan-cancer Classification              & Pan-cancer                                           \\
2     & Golden-standard Subtype Classification & GS-BRCA,~GS-COAD, GS-GBM,~GS-LGG,~GS-OV              \\
3     & Cancer Subtype Clustering              & ACC, KIRP, KIRC, LIHC, LUAD, LUSC, PRAD, THCA, THYM  \\
4     & Omics Data Imputation                  & Imp-BRCA, Imp-COAD, Imp-GBM, Imp-LGG, Imp-OV         \\
\bottomrule
\end{tabular}
}
\end{table}

\subsection{Pan-cancer Classification}

Let \( X^{O} = \{x_1, x_2, ..., x_m\} \) represent the multi-omics dataset, where each \( x_i \) is a vector of features in \( O \)-th omics for the \( i \)-th sample. 
Let \( Y \) denote the set of possible cancer types. 
The goal of cancer classification using multi-omics data is to predict the true label \( y_i \) for each sample \( x_i \) in \( X \), where \( y_i \) belongs to the set of possible cancer types \( Y \). 
Cancer classification can be formulated as a supervised learning problem, where the objective is to learn a mapping function \( f: X \rightarrow Y \) that accurately predicts the true labels for unseen samples based on their omics features.

\subsection{Cancer Subtype Clustering} 

Cancer subtyping means categorizing patients into subgroups that exhibit differences in various aspects based on their multi-omics data.
However, for most cancer types, especially rare cancers, the cancer subtyping tasks are still open questions under discussion.
Thus cancer subtyping tasks are typically clustering tasks without ground true labels.
Let \( X^{O} = \{x_1, x_2, ..., x_m\} \) represent the multi-omics dataset, where each \( x_i \) is a vector of features in \( O \)-th omics for the \( i \)-th sample. 
Let \( k \) denote the set of possible cancer subtypes. 
The goal of cancer subtyping using multi-omics data is to assign each sample \( x_i \) in \( X \) into \( k\) clusters \( C = \{C_1, C_2, ..., C_k\} \), such that each cluster \( C_i \) represents a distinct cancer subtype based on the information from multiple omics data sources.



\subsection{Golden-standard Subtype Classification} 

The cancer research community has thoroughly analyzed the subtypes of some of the most common cancer types in a previous study. Therefore, we consider these subtypes to be the true labels. 
The definition of golden-standard subtype identification is similar to the above Pan-cancer identification tasks.
Golden-standard subtype identification task aims to assign each sample \(x\) in the sample set \(X\) to a cancer subtype \(y\) in the set of all subtypes \(Y\).



\subsection{Omics Data Imputation}

Let \( X \) denote the original omics data with \( m \) samples and \( n \) features, represented as a matrix where \( X_{ij} \) represents the value of the \( i \)-th sample for the \( j \)-th feature. 
Let \( M \) denote the binary mask matrix of the same dimensions as \( X \), where \( M_{ij} = 1 \) if the value of \( X_{ij} \) is observed (not missing), and \( M_{ij} = 0 \) if it is missing. 
The goal of the imputation task is to estimate the missing values in \( X \), denoted as \( \hat{X} \), using the observed values and potentially additional information. 
Imputation can be formulated as \( \hat{X} = f(X, M) \), where \( f \) is the imputation function that takes as input the original omics data \( X \) and the mask matrix \( M \), and outputs the imputed matrix \( \hat{X} \).

\newpage
\section{MLOmics Evaluation Metrics}
\subsection{Precision (Pre)}
Precision measures the accuracy of the positive predictions made by a classification or clustering model. 
It is defined as the ratio of true positive (TP) predictions to the total number of positive predictions made by the model:
\[
Pre = \frac{TP}{TP + FP}
\]
where \(TP\) is the number of true positive predictions (instances correctly classified as positive), and \(FP\) is the number of false positive predictions (instances incorrectly classified as positive).

\subsection{Recall (Re)}  
Recall, also known as sensitivity, measures the ability of a classification or clustering model to identify all relevant instances (i.e., TP) correctly. 
It is defined as the ratio of TP predictions to the total number of actual positive instances:  
\[
Re = \frac{TP}{TP + FN}
\]  
where \(TP\) is the number of true positive predictions (instances correctly classified as positive), and \(FN\) is the number of false negative predictions (instances incorrectly classified as negative).  

\subsection{F1-Score (F1)}  
The F1-score is the harmonic mean of precision and recall, and it provides a balanced measure of a model's accuracy by considering both false positives and false negatives.  
It is particularly useful when the dataset is imbalanced.  
The F1-score is calculated as:  
\[
F1 = 2 \cdot \frac{Pre \cdot Re}{Pre + Re}
\]  
where \(Pre\) is precision, and \(Re\) is recall.  

\subsection{Normalized Mutual Information (NMI)}
Normalized mutual information measures the similarity between two clusterings of the same dataset. 
It measures the mutual dependence between the clustering result and the ground truth labels, normalized by the average entropy of the two clusterings. 
Let \(C\) be the clustering result and \(G\) be the ground truth labels. 
Then, NMI is calculated as:
\[
NMI(C,G) = \frac{I(C,G)}{\sqrt{H(C) \cdot H(G)}}
\]
where \(I(C,G)\) is the mutual information between \(C\) and \(G\),
\(H(C)\) and \(H(G)\) are the entropies of \(C\) and \(G\), respectively.

\subsection{Adjusted Rand Index (ARI)}
Adjusted rand index measures the similarity between two clusterings of the same dataset. 
It measures the agreement between the pairs of samples assigned to the same or different clusters in the two compared clusterings, adjusted for chance. 
ARI is calculated as:
\[
ARI(C,G) = \frac{{a + b}}{{\binom{n}{2}}} - \frac{{a \cdot (a - 1) + b \cdot (b - 1)}}{{\binom{n}{2}}}
\]
where \(a\) is the number of pairs of samples that are in the same cluster in both \(C\) and \(G\), \(b\) is the number of pairs of samples that are in different clusters in both \(C\) and \(G\), \(n\) is the total number of samples, and \(\binom{n}{2}\) is the number of all possible pairs of samples.

\subsection{Silhouette Coefficient (SIL)} 
The silhouette coefficient measures the similarity between a sample and its classified subtype compared to the samples in the other subtypes to determine how appropriately samples in a dataset have been clustered. 
For a sample $i$, let $a(i)$ be the average distance from sample $i$ to other samples in the same cluster, and let $b(i)$ be the smallest average distance from sample $i$ to samples in a different cluster, minimized over clusters. The silhouette coefficient $SIL(i)$ for a sample $i$ is then defined as:
\[
SIL(i) = \frac{b(i) - a(i)}{\max\{a(i), b(i)\}}
\]
The silhouette coefficient ranges from -1 to 1, where a high value indicates that the sample is well-matched to its own cluster and poorly matched to neighboring clusters.

\subsection{P-value of the log-rank Test on Survival Time (LPS)}
The log-rank test on survival time is a hypothesis test used to compare the survival distributions of two or more groups. 
The test statistic $X^2$ is calculated from the observed and expected number of events in each group over time. 
The p-value is then calculated from the test statistic under the null hypothesis that there is no difference in survival distributions between the groups. 
The LPS gives the log-transformed p-values of the log-rank test.
It is calculated as below using the chi-square distribution with $k - 1$ degrees of freedom:
\[
LPS = P(X^2 \geq X^2_{observed})
\]
where $k$ is the number of groups being compared, $X^2_{observed}$ is the observed test statistic calculated from the data.

\subsection{Mean Absolute Error (MAE)}
Mean absolute error measures the average absolute difference between the imputed values and the true values as below:
\[
MAE = \frac{1}{n} \sum_{i=1}^{n} | \hat{Y}_i - Y_i |
\]
where \( n \) is the number of imputed values, \( \hat{Y}_i \) is the imputed value for observation \( i \) and \( Y_i \) is the true value for observation \( i \).

\subsection{Root Mean Squared Error (RMSE)}
Root mean squared error measures the square root of the average squared difference between the imputed values and the true values as below:
\[
RMSE = \sqrt{\frac{1}{n} \sum_{i=1}^{n} (\hat{Y}_i - Y_i)^2}
\]
where \( n \) is the number of imputed values, \( \hat{Y}_i \) is the imputed value for observation \( i \) and \( Y_i \) is the true value for observation \( i \).


\newpage
\section{Downstream Analysis and Biological Resources Linking}
\label{SM: Resources}

\subsection{Differential Gene Expression Analysis}
Differential gene expression analysis has been a cornerstone of transcriptomic studies.
In this analysis, we compare gene expression levels between different experimental conditions or sample groups to identify genes that are significantly upregulated or downregulated. 
Statistical tests such as t-tests or non-parametric tests are commonly used for this purpose. 
For example, gene expression profiles between cancer patients and healthy controls can be compared to identify genes that are dysregulated in cancer. 
Genes with significant differences in expression levels may be further investigated as potential biomarkers or therapeutic targets.
For example, researchers performed differential gene expression analysis on RNA-seq data from Alzheimer's disease patients and healthy controls. 
This analysis identified a panel of differentially expressed genes implicated in neuroinflammation and synaptic dysfunction, showing molecular pathways associated with Alzheimer's disease progression.

We calculated the log2 fold change in gene abundance between pairwise groups and determined the significance of expression changes using Student's t-test. 
P-values were adjusted using the Benjamini-Hochberg procedure to correct the false discovery rate. 
We considered a gene to be significant if it had an adjusted p-value less than 0.05 and a log2 fold change greater than or equal to 1.2. 
Based on their fold changes, the resulting DEGs were categorized into up-regulated and down-regulated sets and can be utilized for subsequent analysis phases.

Among the identified DEGs, several genes have been extensively reported as being associated with cancer progression. 
Notable examples include BRCA1, WNT4, and NOTCH2. 
BRCA1 is well-known for its involvement in hereditary breast cancer and plays essential roles in cell cycle regulation, DNA damage response, and transcriptional control \cite{BRCA1}. 
Dysregulation of the WNT4 gene, which encodes a protein belonging to the Wnt signaling pathway, has been linked to tumor growth, invasion, and metastasis \cite{WNT4}. 
Similarly, the NOTCH2 gene, a member of the Notch receptor family, is critical in cell fate determination, development, and tissue homeostasis and has been implicated in tumor initiation, progression, and therapy resistance \cite{NOTCH2}.

\subsection{Survival Analysis}
Survival analysis is a vital statistical method used to examine and interpret the time until the occurrence of an event, such as death, disease progression, or relapse, in clinical studies. 
It provides insights into factors that influence the survival probability of patients and helps in understanding the impact of clinical, demographic, and molecular variables on patient outcomes.
Common survival analysis techniques include the Kaplan-Meier estimator for survival curves and the Cox proportional hazards model for assessing the relationship between survival and multiple covariates.

In MLOmics, survival analysis is conducted using time-to-event data, such as patient survival time and event status (alive/dead or disease-free/relapsed). 
In our approach, we use log-rank tests to compare survival curves between different groups and assess the significance of survival differences. 
Multivariate Cox regression is employed to evaluate the combined effect of multiple factors on survival. 
Adjustments for confounding variables and interactions are made, and results are presented with hazard ratios and corresponding confidence intervals. 
Survival analysis results are visualized using Kaplan-Meier survival curves.

\subsection{KEGG Pathway Analysis}
Pathway analysis is a critical step in interpreting the biological significance of DEGs. 
By mapping DEGs to known biological pathways, researchers can gain insights into the underlying mechanisms and potential functional impacts of gene expression changes.
In MLOmics, pathway analysis is performed using established databases such as KEGG. 
These databases provide curated information on metabolic pathways, signaling cascades, and gene ontologies. 
DEGs are input into pathway analysis tools to conduct the analysis, which then identifies overrepresented pathways among the upregulated and downregulated gene sets.

For instance, pathway enrichment analysis might reveal that upregulated DEGs in cancer samples are significantly associated with pathways involved in cell cycle regulation and apoptosis, while downregulated DEGs are linked to immune response pathways. Such findings can help to identify potential therapeutic targets and elucidate the molecular basis of disease.

In our approach, we utilize Fisher's exact test or hypergeometric test to evaluate the significance of pathway enrichment. Adjustments for multiple testing are performed using the Benjamini-Hochberg procedure, with pathways considered significant at an adjusted p-value threshold of less than 0.05. 
The pathway analysis results are visualized using enrichment plots and pathway diagrams, which highlight key genes and interactions within the enriched pathways.

\subsection{STRING Network Mapping}
The STRING database \cite{string} aggregates PPIs from experimental data, computational predictions, and curated datasets, offering a standardized framework for network analysis.
STRING network mapping is used to identify and analyze protein-protein interactions (PPIs) among differentially expressed genes. 
This approach facilitates the identification of hub nodes and key interaction pathways in different patients and disease groups.
For example, patient clusters often correspond to functional modules or biological pathways, leading to different gene networks.

In omics analyses, gene identifiers often differ across databases and platforms, which can pose challenges in integrating data for downstream analyses. 
Omics datasets may use Ensembl IDs, Entrez IDs, or gene symbols, while the STRING database requires its own set of identifiers to query PPIs. 
This step is essential for maintaining data consistency and enabling precise network analysis: without proper mapping, some genes might be excluded from the analysis due to identifier mismatches, leading to incomplete or biased results.

In MLOmics, a mapping file resolves these discrepancies by linking MLOmics gene identifiers from omics data to their corresponding STRING identifiers. 
This mapping file is a CSV format file that contains two columns.
The first column is gene identifiers used in the MLOmics dataset.
The second column provides the matching STRING identifiers required for querying the STRING database.
This structure ensures a straightforward lookup for identifier conversion.
Moreover, the CSV format makes inspecting, updating, and adapting this mapping file for other workflows or databases easy. 

Once identifiers are mapped, DEGs can be input into the STRING database to construct interaction networks and further network visualization, typically performed with node attributes (e.g., gene expression values or statistical significance) and edge attributes (e.g., interaction confidence) encoded in the visualization. 

\subsection{Simulate Gene Knockout} 
The simulation begins by ranking all genes based on node degree disparities calculated from the connectivity matrices of the sub-networks. 
Node degree is quantified as the number of direct connections each gene has to other genes within the network, serving as a measure of its centrality and influence across different cancer subtypes.
To derive the connectivity matrices, we analyze the interactions between genes, where each gene is represented as a node and each interaction as an edge. 
The degree of each node is then computed to identify highly interconnected genes.

After ranking, we categorize the genes into two sets: a \emph{high-ranking gene set}, which includes genes exhibiting the largest degree disparities (above a defined threshold based on node degree variance), and a \emph{low-ranking gene set}, composed of genes with minimal degree differences (below the same threshold). 
Using node degree variance as a threshold ensures our classification is statistically grounded.
This method isolates genes that play critical roles in the network dynamics.

Next, we individually simulate the knockout of genes within the high-ranking and low-ranking gene sets. 
This process involves transforming their expression values to a baseline non-expression level, which is defined as either zero or a predefined low expression value (such as the mean expression level of the lowest 10\% of genes). 
This transformation mimics the functional loss of these genes.
For each gene target in the selected sets, we systematically replace its expression value in the patient samples with the baseline non-expression level.

\newpage
\section{Data Source Ethics and Policies}
\label{SM: Ethics}

The ultimate goal of data source ethics and policies was to develop research policies maximizing public benefit from the data that were by these ethical and legal guidelines, ensuring:
(1) Protection of human participants in the project, including their privacy;
(2) Secure and compliant access to TCGA data;
(3) Timely data release to the research community;
(4) Initial scientific publication by the data producers;
(5) These policies have influenced the field of cancer genomics and will continue to serve as a guide for future genomic research projects.

\subsection{Human Subjects Protection and Data Access Policies}
NCI and NHGRI developed a set of policies to protect the privacy of participants donating specimens to TCGA. 
TCGA’s informed consent policy, data access policy, and information about compliance with the HIPAA Privacy Rule are included.

\subsection{Data Use Certification Agreement}
Researchers must agree to A set of policies before gaining access to TCGA data. 
This agreement ensures that researchers pursuing a research question requiring controlled-access data comply with TCGA policies, such as maintaining participants’ privacy, securely accessing the data, and following TCGA publication guidelines. 

\subsection{Suggested Informed Consent Language for Prospective Collections}
An example informed consent document that TCGA suggested Tissue Source Sites use when collecting specimens from prospective project participants. 
This document helps ensure that patients considering donating tissue specimens to human genomics research programs such as TCGA recognize the risks and benefits of participation and understand the nature of their inclusion in the project.

\subsection{Sharing Data from Large-scale Biological Research Projects}
Principles for sharing and publishing genomic data to maximize public benefit developed at a meeting in Fort Lauderdale sponsored by the Wellcome Trust.
These “Fort Lauderdale Principles” informed the original TCGA publication guidelines, which balance making genomic data immediately available for research use with protecting the original owner's initial publication rights. 

\subsection{Considerations for Open Release of Genomic Data from Human Cancer Cell Lines}
An explanation of the factors considered in the decision by NCI and NHGRI to release genomic data and information from the Cancer Cell Line Encyclopedia as open-access data.

\newpage
\section{Limitations \& Broader Impact}
\label{SM: Impact}
This research proposes a benchmark for cancer multi-omics data analysis. 
However, we collected all data from TCGA sources and did not conduct wet experiments to introduce new data further. 
Consequently, the data is limited and influenced by the specific cohorts and methodologies used in TCGA, which may not fully represent the diversity of cancer types or the broader patient population.
Cancer omics data also raises ethical issues, particularly in cancer risk prediction and the development of anti-cancer drugs, which could have potentially harmful or controversial functions. 
The use of omics data for predictive purposes can lead to concerns about privacy, discrimination, and the psychological impact on individuals who are identified as high-risk. 
Additionally, designing drugs based on omics data can lead to unintended side effects and ecological impacts if not carefully regulated.

Nevertheless, we believe that omics data has great potential to benefit society. 
It can lead to more personalized and effective treatments, early cancer detection, and a better understanding of cancer biology. 
Negative impacts can be mitigated through stringent industry regulations, ethical guidelines, and legislation to ensure responsible use and data protection.
The proposed benchmark helps the community develop new cancer omics data analysis algorithms and evaluate the performance of existing models. 
By providing a standardized framework, we aim to facilitate advancements in cancer research and improve the reproducibility and comparability of different computational approaches.

\end{document}